# GPTIPS 2: an open-source software platform for symbolic data mining

**Dominic P. Searson**

School of Computing Science, Newcastle University, UK
(Email: searson@gmail.com)

**Abstract** GPTIPS is a free, open source MATLAB based software platform for symbolic data mining (SDM). It uses a multigene variant of the biologically inspired machine learning method of genetic programming (MGGP) as the engine that drives the automatic model discovery process. Symbolic data mining is the process of extracting hidden, meaningful relationships from data in the form of symbolic equations. In contrast to other data-mining methods, the structural transparency of the generated predictive equations can give new insights into the physical systems or processes that generated the data. Furthermore, this transparency makes the models very easy to deploy outside of MATLAB.

The rationale behind GPTIPS is to reduce the technical barriers to using, understanding, visualising and deploying GP based symbolic models of data, whilst at the same time remaining highly customisable and delivering robust numerical performance for power users. In this chapter, notable new features of the latest version of the software - GPTIPS 2 - are discussed with these aims in mind. Additionally, a simplified variant of the MGGP high level gene crossover mechanism is proposed. It is demonstrated that the new functionality of GPTIPS 2 (a) facilitates the discovery of compact symbolic relationships from data using multiple approaches, e.g. using novel gene-centric visualisation analysis to mitigate horizontal bloat and reduce complexity in multigene symbolic regression models (b) provides numerous methods for visualising the properties of symbolic models (c) emphasises the generation of graphically navigable libraries of models that are optimal in terms of the Pareto trade off surface of model performance and complexity and (d) expedites real world applications by the simple, rapid and robust deployment of symbolic models outside the software environment they were developed in.

# 1 Introduction

Genetic programming (GP; [1]) is a biologically inspired machine learning method that evolves computer programs to perform a task. It does this by randomly generating a population of computer programs (usually represented by tree structures) and then breeding together the best performing trees to create a new population. Mimicking Darwinian evolution, this process is iterated until the population contains programs that solve the task well.

When building an empirical mathematical model of data acquired from a process or system, the process is known as symbolic data mining (SDM). SDM is an umbrella term to describe a variety of related activities including generating symbolic equations predicting a continuous valued response variable using input/predictor variables (symbolic regression); predicting the discrete category of a response variable using input variables (symbolic classification, e.g. see [2,3]) and generating equations that optimise some other criterion (symbolic optimisation, e.g. GPTIPS was used in this way to generate new chaotic attractors in [4]).

Symbolic regression is perhaps the most well known of these activities (it is closely related to classical regression modelling) and the most widely used. Hence, much of the functionality of GPTIPS is targeted at facilitating it. Unlike traditional regression analysis (in which the user must specify the structure of the model and then estimate the parameters from the data), symbolic regression automatically evolves both the structure and the parameters of the mathematical model from the data. This allows it to both select the inputs (features) of the model and capture non-linear behaviour.

Symbolic regression models are typically of the form:

$$\hat{y} = f(x_1, ..., x_M) \tag{1}$$

where $y$ is an output/response variable (the variable/property you are trying to predict), $\hat{y}$ is the model prediction of $y$ and $x_1, ..., x_M$ are input/predictor variables (the variables/properties you know and want to use to predict $y$; they may or may not in fact be related to $y$) and $f$ is a symbolic non-linear function (or a collection of non-linear functions). A typical simple symbolic regression model is:

$$\hat{y} = 0.23\,x_1 + 0.33(x_1 - x_5) + 1.23\,x_3{}^2 - 3.34\,\cos(x_1) + 0.22 \tag{2}$$

This model contains both linear and non-linear terms and the structure and parameterisation of these terms is automatically determined by the symbolic regression algorithm. Hence, it can be seen that symbolic regression provides a flexible – yet simple – approach to non-linear predictive modelling.

Additional advantages of symbolic regression are:

- It can automatically create compact, accurate equations to predict the behaviour of physical systems. This appeals to the notion of Occam's razor. In particular, the use of multigene GP (MGGP) within GPTIPS can exert a 'remarkable' degree of control of model complexity in comparison with standard GP [5].

- Unlike many soft-computing modelling methodologies - such as feed forward artificial neural networks or support vector machines (SVMs) - no specialised modelling software environment is required to deploy the trained symbolic models. And, because the symbolic models are simple constitutive equations, a non-modelling expert can easily and rapidly implement them in any modern computing language. Furthermore, the simplicity of the model form means they are more maintainable than typical black box predictive models.

- Examination of the evolved equations can often lead to human insight into the underlying physical processes or dynamics. In addition, the ability of a human user to understand the terms of a predictive equation can help instil trust in the model [6]. It is hard to overstate the importance of user understanding and trust in predictive models, although this is not often discussed in the predictive modelling literature. In contrast, it is extremely difficult, if not impossible, to gain insight into a neural net model where the 'knowledge' about the data, system or process is encoded as network weights.

- Discovery of a population of models (rather than a single model as in the majority of other predictive modelling techniques). The evolved population can be regarded as a model library and usually contains diverse models of varying complexity and performance. This gives the user choice and the ability to gain understanding of the system being modelled by examination of the model library.

Note that the human related factors mentioned above, such as interpretation and deployment of models, are especially important when dealing with data obtained from highly multivariate non-linear systems of unknown structure [6] for which traditional analysis tends to be difficult or intractable.

Hence, symbolic regression (and symbolic data mining in general) has many features that make it an attractive basis for inducing simple, interpretable and deployable models from data where the 'true' underlying relationships are high dimensional and largely unknown. However, there has been a relative paucity of software that allows researchers to actually do symbolic data mining, and in many cases the existing software is either expensive, proprietary and closed source or requires a high degree of expertise in software configuration and machine learning to use it effectively.

GPTIPS (an acronym for Genetic Programming Toolbox for the Identification of Physical Systems) was written to reduce the technical barriers to using symbolic data mining and to help researchers, who are not necessarily experts in computing science or machine learning, to build and deploy symbolic models in their fields of research. It was also written to promote understanding of the model discovery mechanisms of MGGP and to allow researchers to add their own custom implementations of code to use MGGP in other non-regression contexts (e.g. [4]). To this end, it was written as a free (subject to the GNU public software license, GPL v3), open source project in MATLAB.

The use of MATLAB as the underlying platform confers the following benefits:

- Robust, trustable, fast and automatically multi-threaded implementations of many matrix and vector math algorithms (these are used extensively in GPTIPS).
- Widely taught at the undergraduate level and beyond at educational institutes around the world and hence is familiar (and site licensed) to a diverse array of students, researchers and other technical professionals. It is also heavily used in many commercial, technical and engineering environments.
- Supported, regularly updated and bug fixed and extremely well documented.
- Easy to use interface and interactive environment and supports the import and export of data in a wide variety of formats.
- A robust symbolic math engine (MuPAD) that is exceptionally useful for the post-run processing, simplification, visualisation and export of symbolic models in different formats using variable precision arithmetic.
- Runs on many OS platforms (i.e. Windows, Linux, Mac OSX) using the same code.
- Increasing emphasis on parallel computing (e.g. GPTIPS 2 has a parallel mode and can use unlimited multiple cores to evolve and evaluate new models), GPU computing, cloud computing and other so called 'big data' features such as memory-mapped variables.

This chapter is structured as follows: Section 2 provides a high level overview of GPTIPS and, in particular, the new features aimed at multigene regression model development in GPTIPS2. Section 3 is provided to review some different forms of symbolic regression in the context of classical regression analysis and describes the mechanisms of MGGP. Note that a basic tutorial level description of 'standard' GP is not provided here, as it is readily available elsewhere, e.g. [7]. Section 4 is used to demonstrate some of the features of GPTIPS 2, focusing on the visual analytics tools provided for the development of portable multigene symbolic regression models. Section 5 describes a new gene-centric approach to identifying and removing horizontal bloat in multigene regression models, with emphasis on the new visual analysis tool provided in GPTIPS to do this. Finally, the chapter ends with some concluding remarks in Section 6.

## 2 GPTIPS 2 – Overview

GPTIPS (version 1) has become a widely used technology platform for symbolic data mining via MGGP. It is used by researchers globally and has been successfully deployed in dozens of application areas. [1]

GPTIPS using MGGP based regression has been shown to outperform existing soft-computing/machine learning methods such as neural networks, support vector machines etc. on many problem domains in terms of predictive performance and model simplicity. Examples include:

- Global solar irradiation prediction – MGGP was noted to give clearly better results than fuzzy logic and neural networks and the resulting equations were understandable by humans [8].
- The automated derivation of correlations governing the fundamental properties of the motion of particles in fluids, a key subject in powder technology, chemical and environmental engineering. The evolved models were significantly better (up to 70%) than the existing empirical correlations [9].
- The reverse engineering of the structure of the interactions in biological transcription networks from time series data, attaining model accuracy of around 99% [10].
- The use of MGGP for the accurate modelling and analysis of data from complex geotechnical and earthquake engineering problems [5, 11]. It was noted that the evolved equations were highly accurate and 'particularly valuable for pre-design practices' [5].

The symbolic engine of GPTIPS, i.e. the mechanism whereby new equations are generated and improved over a number of iterations, is a variant of GP called multigene genetic programming (MGGP, e.g. see [12, 13, 14]) which uses a modified GP algorithm to evolve data structures that contain multiple trees (genes). An example of a single tree representing a gene is shown in Fig. 1. This represents the equation $\sin(x_1) + \sin(3x_1)$. A typical GPTIPS multigene regression model consists of a weighted linear combination of genes such as these.

---

[1] A list of research literature using GPTIPS is maintained at
https://sites.google.com/site/gptips4matlab/application-areas

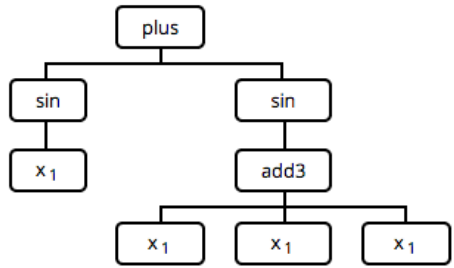

**Fig. 1.** Example of a tree (gene) representing the model term $\sin(x_1) + \sin(3x_1)$. This tree visualisation was created as a graphic within an HTML file using the GPTIPS 2 `drawtrees` function. The appearance of the trees is user customisable using simple CSS.

GPTIPS is a generic tree based GP platform and has a pluggable architecture. This means that users can easily write their objective/fitness functions (e.g. for symbolic classification and symbolic optimisation) and plug them into GPTIPS without having to modify any GPTIPS code.

GPTIPS also has many features aimed specifically at developing multigene symbolic regression models. This combines the ability to evolve new equation model terms of MGGP with the power of classical linear least squares parameter estimation to optimally combine these model terms in order to minimise a prediction error metric over a data set. It is sometimes helpful to think of GPTIPS multigene regression models as pseudo-linear models in that they are linear combinations of low order non-linear transformations of the input variables. These transformations can be regarded as meta-variables in their own right.

Multigene symbolic regression has been shown to be able to evolve compact, accurate models and perform automatic feature selection even when there are more than 1500 input variables [14]. It has been demonstrated that multigene symbolic regression can be more accurate and efficient than 'standard' GP for modelling non-linear problems (e.g. see [5, 11]).

## 2.1 GPTIPS Feature Overview

GPTIPS is mostly a command line driven modelling environment and it requires only a basic working knowledge of MATLAB. The user creates a simple configuration file where the data is loaded from file (or generated algorithmically within the configuration file) and configuration options set (numerous example configuration files and several example data sets are provided with GPTIPS). GPTIPS automatically generates default values for the majority of configuration options and these can be modified in the configuration file. Typical configuration options that the user sets are population size, maximum number of generations to run for, number of genes and tournament size. However, there are a large number of other run configuration options that the user can explore. In addition, GPTIPS 2 has the following features to support effective non-linear symbolic model development, analytics, export and deployment:

- Automatic support for the Parallel Computing Toolbox: fitness and complexity calculations are split across multiple cores allowing significant run speedup.
- Automatic support for training, validation and test data sets and comprehensive reporting of performance stats for each.
- An extensive set of functions for tree building blocks is provided: plus, minus, multiply, divide (protected and unprotected), add3 (ternary addition), mult3 (ternary multiplication), tanh, cos, sin, exp, $\log_{10}$, square, power, abs, cube, sqrt, exp (- x), if-then-else, -x, greater than (>), less than (<), Gaussian (exp ($x^2$)) and threshold and step functions. Furthermore – virtually any built in MATLAB math function can be used a tree building block function (sometimes a minor modification is required such as writing a wrapper function for the built in function). In general, it is very easy for users to define their own building block functions.
- Tight integration with MATLAB's MuPAD symbolic math engine to facilitate the post-run analysis, simplification and deployment of models.
- Run termination criteria. In addition to number of generations to run for, it is usually helpful to specify additional run termination criteria in order to avoid waste of computational effort. In GPTIPS, the maximum amount of time to run for (in seconds) can be set for each run as well as a target fitness. E.g. for multigene regression the target fitness can be set as model root mean squared error (RMSE) on the training data.
- Multiple independent runs where the populations are automatically merged after the completion of the runs. It is usually beneficial to allocate a relatively small amount of computational effort to each of multiple runs rather than to perform a single large run (e.g. 10 runs of 10 seconds each rather than a single run of 100 seconds). E.g. this 'multi-start' approach mitigates problems with the possible loss of model diversity over a run and with the GP algorithm getting stuck in local minima. In addition, GPTIPS 2 provides functionality such that final populations of separate runs may be manually merged by the user.
- Steady-state GP and fitness caching.
- Two measures of tree complexity: node count and expressional complexity [6]. The latter is a more fine-grained measure of model complexity and is used to promote flatter trees over deep trees. This has significant benefits (albeit at extra computation cost) in evolving compact, low complexity models. For a single tree, expressional complexity is computed by summing together the node count of itself and all its possible *full* sub-trees (a leaf node is also considered a full sub-tree) as illustrated in [6]. Hence, for two trees with the same node count, flatter and balanced trees have a lower expressional complexity than deeper ones. For instance, the tree shown in Fig. 1 has a total node count of 8 and contains 8 possible sub-trees. The sum of the node counts of the 8 possible full sub-trees gives, in this case, an expressional complexity of 23. For multigene individuals, the overall expressional complexity is computed as the simple sum of the expressional complexities of its constituent trees.

- Regular tournament selection (considers fitness only), Pareto tournament selection (considers fitness and model complexity) and lexicographic tournament selection (similar to regular tournament selection but always chooses the less complex model in the event of a fitness 'tie'). The user can set the probability of a particular tournament type occurring at every selection event (i.e. each time the GP algorithm selects an individual for crossover, mutation etc.). E.g. the user can set half of all selection events to be performed by regular tournament and half by Pareto tournament. Pareto tournaments of size $P$ for two objectives are implemented using the $O(P^2)$ fast non-dominated sort algorithm described in [15][2].

- 6 different tree mutation operators.

- Interactive graphical population browser showing Pareto front individuals in terms of fitness (or for multigene regression models, the coefficient of determination $R^2$) and complexity on training, validation and test data sets. This facilitates the exploration of multigene regression models that are accurate but not overly complex and the identification of models that generalise well across data sets.

- A configurable multigene regression model filter object that enables the progressive refinement of populations according to model performance, model complexity and other user criteria (e.g. the presence of certain input variables in a model).

- Functions to export any symbolic regression model to (a) a symbolic math object (b) a standalone MATLAB file for use outside GPTIPS (c) snippets of optimised C code – which may be easily manually ported to other languages such as Java (d) an anonymous MATLAB function or function handle (e) an HTML formatted equation (f) a LaTeX formatted equation (g) a MATLAB data structure containing highly detailed information on the model as well as the individual gene predictions on training, test and validation data.

- Standalone (i.e. can be viewed in a web browser without the need for MATLAB) HTML model report generator. This enables a comprehensive performance and statistical analysis of any model in the population to be exported to HTML for later reference. The HTML report contains interactive graphical displays of model performance and model genotype and phenotype structure.

---

[2] Currently, the Pareto tournament implementation does not support more than 2 objectives.

- Customisable standalone HTML model report generator to visualise the tree structure(s) comprising an individual/model.
- Standalone HTML Pareto front report generator to allow the interactive visualisation of simplified multigene regression models in tabular format, sortable by performance (in terms of the coefficient of determination, i.e. model $R^2$) and model complexity.
- Regression Error Characteristic (REC; [16]) curves to allow simple graphical comparisons of the predictive performance of selected multigene regression models.

# 3 Multigene Symbolic Regression and MGGP – Overview and Mathematical context

In this section, multigene symbolic regression is described in a mathematical context and compared with some other common symbolic regression methods as well as multiple linear regression (MLR). In addition, the mechanics of the MGGP algorithm are described, including a new, simplified high level crossover operator to expedite the exchange of genes between individuals during the simulated evolutionary process.

## *3.1 Multigene Symbolic Regression*

### 3.1.1 Naïve Symbolic Regression

In early standard formulations of symbolic regression (which will be referred to as naïve symbolic regression) GP was often used to evolve a population of trees, each of which is interpreted *directly* as a symbolic mathematical equation that predicts a ($N \times 1$) vector of outputs/responses **y** where $N$ is the number of observations of the response variable $y$. The corresponding input matrix **X** is an ($N \times M$) data matrix where $M$ is the number of input variables. In general, only a subset of the $M$ variables are 'selected' by GP to form the models. In naïve symbolic regression, the $i$th column of **X** comprises the $N$ input values for the $i$th variable and is designated the input variable $x_i$. Fig. 2 illustrates naïve symbolic regression.

Typically, the GP algorithm will attempt to minimise the sum of squared errors (SSE) between the observed response **y** and the predicted response $\hat{\mathbf{y}}$ (where the ($N \times 1$) error vector **e** is **y** - $\hat{\mathbf{y}}$) although other error measures are also frequently used, e.g. the mean squared error (MSE) and the root mean squared error (RMSE), the latter having the advantage that it is expressed in the units of the response variable $y$.

$\hat{\mathbf{y}} =$ 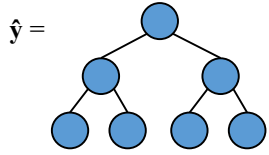

**Fig. 2.** Naïve symbolic regression. The prediction of the response data **y** is the unmodified output of a single tree that takes as its inputs one or more columns of the data matrix **X**.

### 3.1.2 Scaled Symbolic Regression

To improve the efficacy of symbolic regression a bias (offset) term $b_0$ and a weighting/scaling term $b_1$ can be used to modify the tree output so that it fits **y** better. The values of these coefficients are determined by linear least squares and, for any valid tree, the prediction is guaranteed to be at least as good as the naïve prediction. It will almost always be better (the only case where it is not is the case $b_0 = 0$ and $b_1 = 1$). This method is essentially the same as scaled symbolic regression [17] because the coefficients $b_0$ and $b_1$ translate and linearly scale the raw output of the tree in such a way as to minimise the prediction error of **y** as shown in Fig. 3.

$\hat{\mathbf{y}} = b_0 + b_1 \times$ 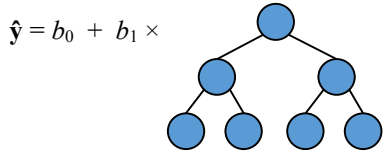

**Fig. 3.** Scaled symbolic regression. The prediction of the response data **y** is the vector output of single tree modified by a bias term $b_0$ and a scaling parameter $b_1$. These are determined by linear least squares.

Hence, the prediction of **y** is given by:

$$\hat{\mathbf{y}} = b_0 + b_1 \, \mathbf{t} \qquad (3)$$

where **t** is the $(N \times 1)$ vector of outputs from the GP tree on the training data. This may also be written as:

$$\hat{\mathbf{y}} = \mathbf{Db} \qquad (4)$$

where **b** is a $(2 \times 1)$ vector comprising the $b_0$ and $b_1$ coefficients and **D** is an $(N \times 2)$ matrix where the 1st column is a column of ones (this is used as a bias/offset input) and the 2nd column is the tree outputs **t**. The optimal linear least squares estimate (i.e. that which minimises the SSE $\mathbf{e}^T\mathbf{e}$) of **b** is computed from **y** and **D** using the well known least squares normal equation as shown in (5) where $\mathbf{D}^T$ is the matrix transpose of **D**. Note that the optimality of the estimate of **b** is only strictly true if a number of assumptions are met such as independence of the columns of **D** and normally distributed errors. In practice, these assumptions are rarely strictly met – but with the use of the Moore-Penrose pseudo-inverse (described in the following section) – the violations of these assumptions do not appear to prevent the practical development of effective symbolic regression models.

$$\mathbf{b} = (\mathbf{D}^{\mathrm{T}}\mathbf{D})^{-1}\mathbf{D}^{\mathrm{T}}\mathbf{y} \tag{5}$$

### 3.1.3 Multigene Symbolic Regression

A generalisation of the previous approach is to use $G$ trees to predict the response data **y**. GPTIPS uses MGGP to evolve the trees comprising the additive model terms in each individual and this is referred to as multigene symbolic regression.

Again, there is an offset/bias coefficient $b_0$ and now the coefficients $b_1$, $b_2$, ..., $b_G$ are used for scaling the output of each tree/gene. A linear combination of scaled tree outputs can capture non-linear behaviour much more effectively than using scaled symbolic regression, in which one tree must capture all of the non-linear behaviour.

Moreover, by enforcing depth restricted trees and using other strategies such as Pareto tournaments and expressional complexity, this leads to the evolution of compact models that tend to have linearly separable terms and so lend themselves to automated post-run model simplification using symbolic math software. The structure of multigene symbolic regression models is illustrated in Fig. 4.

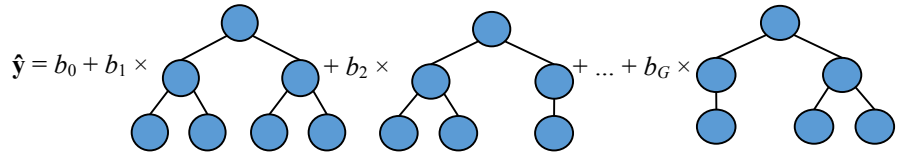

**Fig. 4.** Multigene symbolic regression. The prediction of the response data **y** is the vector output of $G$ trees modified by bias term $b_0$ and scaling parameters $b_1$, ..., $b_G$.

The prediction of the **y** training data is given by:

$$\hat{\mathbf{y}} = b_0 + b_1\,\mathbf{t}_1 + \ldots + b_G\,\mathbf{t}_G \tag{6}$$

where $\mathbf{t}_i$ is the $(N \times 1)$ vector of outputs from the $i$th tree/gene comprising a multigene individual. Next, define **G** as a $(N \times (G + 1))$ gene response matrix as follows in (7).

$$\mathbf{G} = [\mathbf{1}\ \mathbf{t}_1\ \ldots\ \mathbf{t}_G] \tag{7}$$

where the **1** refers to a ($N \times 1$) column of ones used as a bias/offset input.

Now (6) can be rewritten as:

$$\hat{\mathbf{y}} = \mathbf{Gb} \tag{8}$$

The least squares estimate of the coefficients $b_0$, $b_1$, $b_2$,..., $b_G$ formulated as a (($G + 1) \times 1$) vector can be computed from the training data as:

$$\mathbf{b} = (\mathbf{G}^T \mathbf{G})^{-1} \mathbf{G}^T \mathbf{y} \tag{9}$$

In practice, the columns of the gene response matrix **G** may be collinear (e.g. due to duplicate genes in an individual, and so the Moore-Penrose pseudo-inverse (by means of the singular value decomposition; SVD) is used in (9) instead of the standard matrix inverse. Because this is computed for every individual in a GPTIPS population at each generation (except for cached individuals), the computation of the gene weighting coefficients represents a significant proportion of the computational expense of a run. In GPTIPS, the RMSE is then calculated from $\mathbf{e}^T\mathbf{e}$ and is used as the fitness/objective function that is minimised by the MGGP algorithm[3].

Compare this with classical MLR which is typically of the form:

$$\hat{\mathbf{y}} = a_0 + a_1 \mathbf{x}_1 + a_2 \mathbf{x}_2 + ... + a_N \mathbf{x}_M \tag{10}$$

Here, the data/design matrix **X** is defined as:

$$\mathbf{X} = [\mathbf{1} \ \mathbf{x}_1 \ ... \ \mathbf{x}_M] \tag{11}$$

and this allows the least squares computation of the coefficients $a_0$, $a_1$, ... $a_M$ as:

$$\mathbf{a} = (\mathbf{X}^T \mathbf{X})^{-1} \mathbf{X}^T \mathbf{y} \tag{12}$$

where **a** is a (($M + 1) \times 1$) vector containing the $a$ coefficients.

This section described how a multigene individual can be interpreted as a linear-in-the-parameters regression model and how the model coefficients are computed using least squares. The following section outlines how MGGP actually generates and evolves the trees that the form the component genes of multigene regression models.

---

[3] Although RMSE is the default fitness measure, this can be easily changed to, for example, MSE by a very minor edit to the file containing the default fitness function.

### *3.2 Multigene Genetic Programming*

Here it is outlined how multigene individuals are created and then iteratively evolved by the MGGP algorithm. This algorithm is similar to a 'standard' GP algorithm except for modifications made to facilitate the crossover and mutation of multigene individuals. Note that - although GPTIPS uses MGGP primarily for symbolic regression - the algorithmic implementation of MGGP is *independent* of the interpretation of the multigene individuals as regression models. Multigene individuals can also be used in other contexts, e.g. classification trees [3]. In GPTIPS there is a clear modular separation of the MGGP code and the code that implements multigene regression. GPTIPS has a simple pluggable architecture in that it provides explicit code hooks to allow the addition of new code that interprets multigene individuals in a way of the user's choosing (the code for performing multigene regression is - by default - attached to these hooks). Note that MGGP also implicitly assumes that the specific ordering of genes in any individual is unimportant.

In the first generation of the MGGP algorithm, a population of random individuals is generated (it is currently not possible to seed the population with partial solutions). For each new individual, a tree representing each gene is randomly generated (subject to depth constraints) using the user's specified palette of building block functions and the available $M$ input variables $x_1, \ldots, x_M$ as well as (optionally) ephemeral random constants (ERCs) which are generated in a range specified by the user (the default range is -10 to 10). ). In the first generation the MGGP algorithm attempts to maximise diversity by ensuring that no individuals contain duplicate genes. However, due to computational expense, this is not enforced for subsequent generations of evolved individuals.

Each individual is specified to contain (randomly) between 1 and $G_{max}$ genes. $G_{max}$ is a parameter set by the user. When using MGGP for regression, a high $G_{max}$ may capture more non-linear behaviour but there is the risk of overfitting the training data and creating models that contain complex terms that contribute little or nothing to the model's predictive performance (horizontal bloat). This is discussed further in Section 5. Conversely, setting $G_{max}$ to 1 is equivalent to performing scaled symbolic regression.

As in standard GP, at each generation individuals are selected probabilistically for breeding (using regular or Pareto tournaments or a mixture of both). Each tournament results in an individual being selected based on either its fitness or – for Pareto tournaments - its fitness and its complexity (the user can set this to be either the total node count of all the genes in an individual or the total expressional complexity of all the genes in an individual).

In MGGP, there are two types of crossover operators: high level crossover and the standard GP sub-tree crossover, which is referred to as low level crossover. The high level crossover operator is used as a probabilistically selected alternative to the ordinary low level crossover (in GPTIPS the default is that approximately a fifth of crossover events are high level crossovers).

When low level crossover is selected a gene is randomly chosen from each parent. These genes undergo GP sub-tree crossover with each other and the offspring genes replace the original genes in the parent models. The offspring are then copied into the new population.

When high level crossover is selected an individual may acquire whole genes - or have them deleted. This allows individuals to exchange one or more genes with another selected individual (subject to the $G_{max}$ constraint).

In GPTIPS 2 the high level crossover operator described in [12, 13, 14] has been simplified and is outlined below between a parent individual consisting of the 3 genes labelled (G1 G2 G3) and a parent individual consisting of the genes labelled (G4 G5 G6 G7) where (in this hypothetical case) $G_{max}$ = 5.

**Parents**:　　　(G1 **G2** G3)
　　　　　　　　(**G4** G5 G6 **G7**)

A crossover rate parameter *CR* (where $0 < CR < 1$) is defined. This is similar to the CR parameter used in differential evolution (DE, see [18]) and a uniform random number *r* between 0 and 1 is generated independently for each gene in the parents. If r is $\leq CR$ then the corresponding gene is moved to the other individual. The default value of *CR* in GPTIPS 2 is 0.5.

Hence, randomly selected genes (highlighted in boldface above) are exchanged resulting in two offspring in the next generation.

**Offspring**:　　　(G1 G3 **G4 G7**)
　　　　　　　　(G5 G6 **G2**)

This high level crossover mechanism is referred to as *rate based high level crossover* to distinguish it from the *two point high level crossover* mechanism in GPTIPS version 1 (which swapped contiguous sections of genes from individuals). Note that the rate based high level crossover mechanism results in new genes for both individuals as well as reducing the overall number of genes for one model and increasing the total number of genes for the other. If an exchange of genes results in either offspring containing more genes than the $G_{max}$ constraint then genes are randomly deleted until the constraint is no longer violated.

# 4 Using GPTIPS

In this section it will be illustrated how GPTIPS 2 may be used to generate, analyse and export non-linear multigene regression models, both using command line tools and visual analytics tools and reports. The example screenshots in the Figures contained in this section are taken from example runs from various data sets using configuration files and data that are provided with GPTIPS 2. The screenshots were obtained using MATLAB Release 2014b on OSX.

## 4.1 Running GPTIPS

As discussed in Section 2.1, the user creates a simple text configuration file that specifies some basic run parameters and either loads in the data to be modelled from file or algorithmically generates it. Any unspecified parameters are set to GPTIPS default values.

To run the configuration file (here called `configFileName.m`) the `rungp` function is used as follows:

```
gp = rungp(@configFileName)
```

where the @ symbol denotes a MATLAB function handle to the configuration file.

The GPTIPS run then begins. When it is complete – the population and all other relevant data is stored in the MATLAB 'struct' variable `gp`. This is used as a basis for all subsequent analyses.

## 4.2 Exploratory Post Run Analyses

GPTIPS provides a number of exploratory post-run interactive visualisation and analysis tools. For instance, a simple summary of any run can be generated using the `summary` function and an example is shown in Fig. 5.

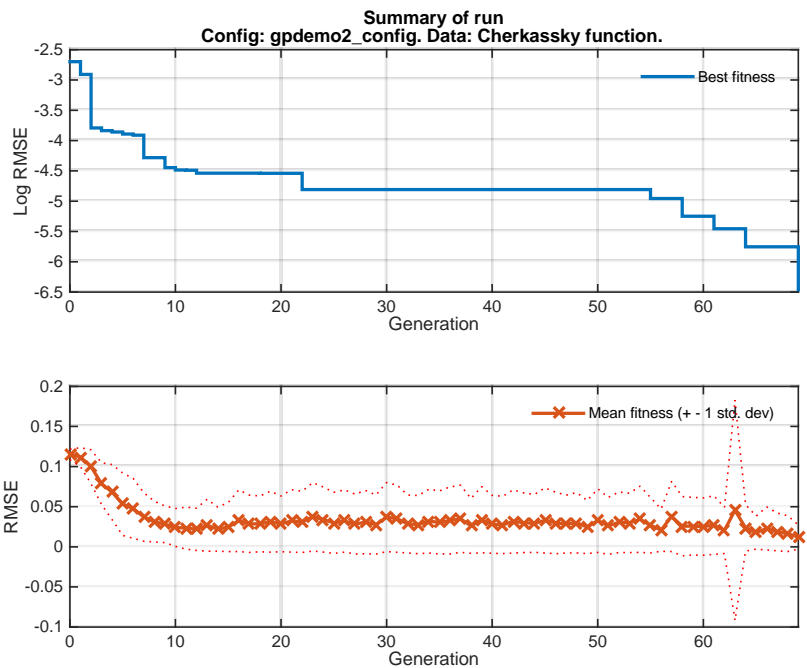

**Fig. 5.** An example of a run summary in GPTIPS. Generated using the `summary` function.

For multigene symbolic regression this shows in the upper part of the chart – by default – the $\log_{10}$ value of the best RMSE (this is the error metric that GPTIPS attempts to minimise over the training data) achieved in the population over the generations of a run. The lower part of the chart shows the mean RMSE achieved in the population.

Other tools are intended to help the user to identify a model (or small set of models) that look promising and worthy of further investigation. One of the most useful visual analytic tools is the population browser. This interactive tool visually illustrates the entire population in terms of its predictive performance and model complexity characteristics. This is generated using the `popbrowser` function. An example of this is shown in Fig. 6. Each model is plotted as a dot with (1- $R^2$) on the vertical axis and expressional complexity on the horizontal axis. The Pareto front models are highlighted in green and it is almost always these models that will be of the greatest interest to the user. In particular, the Pareto models in the lower left of the population (high $R^2$ and low complexity) are usually where a satisfactory solution may be found.

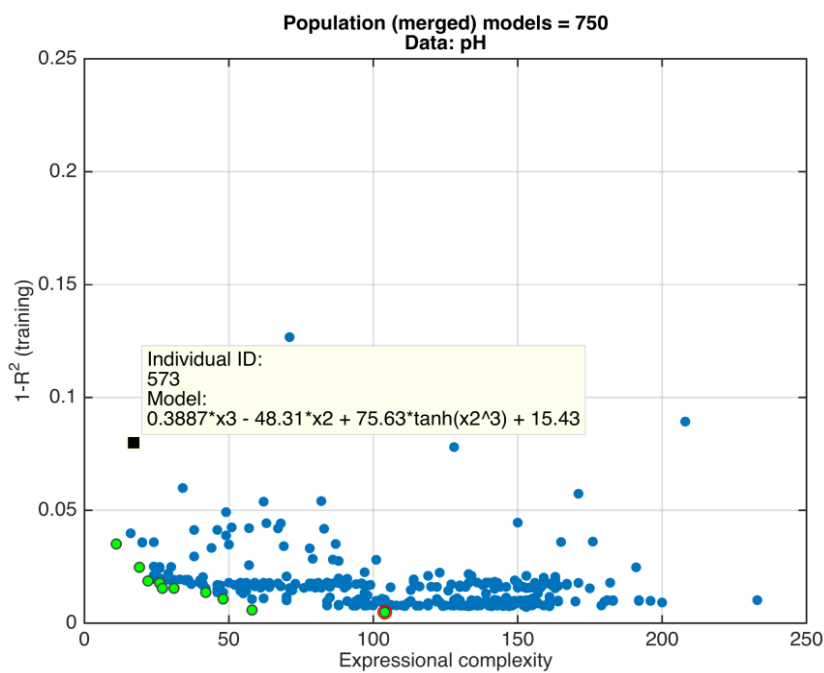

**Fig. 6.** Visually browsing a multigene regression model population. Green dots represent the Pareto front of models in terms of model performance ($1 − R^2$) and model complexity. Blue dots represent non-Pareto models. The red circled dot represents the best model in the population in terms of $R^2$ on the training data. Clicking on a dot shows a yellow popup containing the model ID and the simplified model equation. Generated using the `popbrowser` function.

This visualisation may be used with the training, validation or test data sets. For example Fig. 6 was generated using:

```
popbrowser(gp,'train')
```

Another way of displaying information about Pareto front models in a population is by use of the `paretoreport` function. This creates a standalone HTML file – viewable in a web browser – that includes a table listing the simplified model equations along with the model performance and expressional complexity. The table is interactive and the models can be sorted by performance or complexity by clicking on the appropriate column header. An example of an extract from such a report is shown in Fig. 7. This report assists the user in rapidly identifying the most promising model or models to investigate in more detail.

| Model ID | Goodness of fit ($R^2$) | Model complexity | Model |
|---|---|---|---|
| 66 | 0.993 | 49 | $0.00336\, x_3\, x_4\, \tanh(x_1 - 1.0\, x_3) - 0.134\, x_4 - 2.89\, \tanh(x_1 - 1.0\, x_3) - 2.06\text{e-}4\, x_3\, x_4{}^2 - 0.845\, x_1 + 0.00113\, x_1\, x_3\, x_4 + 20.4$ |
| 13 | 0.993 | 37 | $2.63\, x_2 - 0.695\, x_1 + 0.0679\, x_3 - 3.35\, \tanh(x_1 - 1.0\, x_3) + 0.00391\, x_3\, x_4\, \tanh(x_1 - 1.0\, x_3) + 2.65\text{e-}4\, x_1\, x_3\, x_4 + 15.0$ |
| 12 | 0.992 | 26 | $6.53\, x_2 - 1.62\, x_1 - 0.262\, x_3 - 0.972\, \tanh(x_1 - 1.0\, x_3) - 4.92\text{e-}4\, x_3\, x_4{}^2 + 0.00204\, x_1\, x_3\, x_4 + 26.1$ |
| 5 | 0.989 | 18 | $5.0\, x_2 - 0.734\, x_1 + 0.557\, x_3 - 1.22\, \tanh(x_1 - 1.0\, x_3) - 0.00968\, x_3{}^2 + 10.9$ |
| 27 | 0.983 | 17 | $0.828\, x_3 - 1.2\, \tanh(x_1 - 1.0\, x_3) - 7.0\text{e-}4\, x_1\, x_3\, x_4 + 0.685$ |
| 7 | 0.974 | 12 | $14.9\, x_2 + 0.963\, x_3 - 0.788\, x_4 - 1.45\, \tanh(x_1 - 1.0\, x_3) + 10.3$ |
| 36 | 0.964 | 11 | $0.752\, x_4 - 1.79\, x_1 - 1.57\text{e-}4\, x_3\, x_4{}^2 + 15.2$ |
| 49 | 0.953 | 9 | $1.1\, x_3 - 1.56\, x_2 - 8.47\text{e-}4\, x_1\, x_3\, x_4 - 1.55$ |
| 31 | 0.916 | 3 | $8.86\, x_2 + 0.372\, x_3 - 3.29$ |

**Fig. 7.** Extract from a Pareto front model HTML report. GPTIPS 2 can generate a standalone interactive HTML report listing the multigene regression models on the Pareto front in terms of their simplified equation structure, expressional complexity and performance on the training data ($R^2$). The above table is sortable by clicking on the appropriate column header. Generated using the `paretoreport` function.

It is also possible to filter populations according to various user criteria using the `gpmodelfilter` object. The output of this filter is another `gp` data structure which is functionally identical to the original (in the sense that any of the command line and visual analysis tools may be applied to it) except that models not fulfilling user criteria have been removed.

For example, if the user wants to only retain models that (a) have an $R^2$ greater than 0.8 (b) contain the input variables $x_1$ and $x_2$ and (c) do not contain the variable $x_4$ then the filter can be configured and executed as follows:

Create a new filter object `f`:

```
f = gpmodelfilter
```

Next set the user criteria, i.e. models must have $R^2$ (training data) greater or equal to 0.8:

```
f.minR2train = 0.8
```

Must include $x_1$ and $x_2$:

```
f.includeVars = [1 2]
```

Must exclude x4:

```
f.excludeVars = 4
```

Finally, apply the filter to the existing population structure `gp` to create a new one `gpf`:

```
gpf = f.applyFilter(gp)
```

At this point the user may apply the exploratory tools (e.g. `paretoreport`) to the refined population to zero in on models of interest fulfilling certain criteria. Other criteria that can be set include maximum expressional complexity, maximum and minimum number of variables and Pareto front (i.e. exclude all models not on the Pareto front).

## 4.3 Model Performance Analyses

Once a model (or set of models) has been identified using the tools described above, the detailed performance of the model can be assessed by use of the `runtree` function. This essentially re-runs the model on the training data (and validation and test data, if present) and generates a set of graphs including predicted vs actual $y$ and scatterplots of predicted vs actual $y$. These graphs can be generated using the numeric model ID (e.g. from the `popbrowser` visualisation) as an input argument to `runtree` or by using keywords such as 'best' (best model on training data) and 'testbest' (best model on test data), e.g.

```
runtree(gp,'testbest')
```

This is a common design pattern across a large number of GPTIPS functions. An example of the scatterplots generated by `runtree` is shown in Fig. 8.

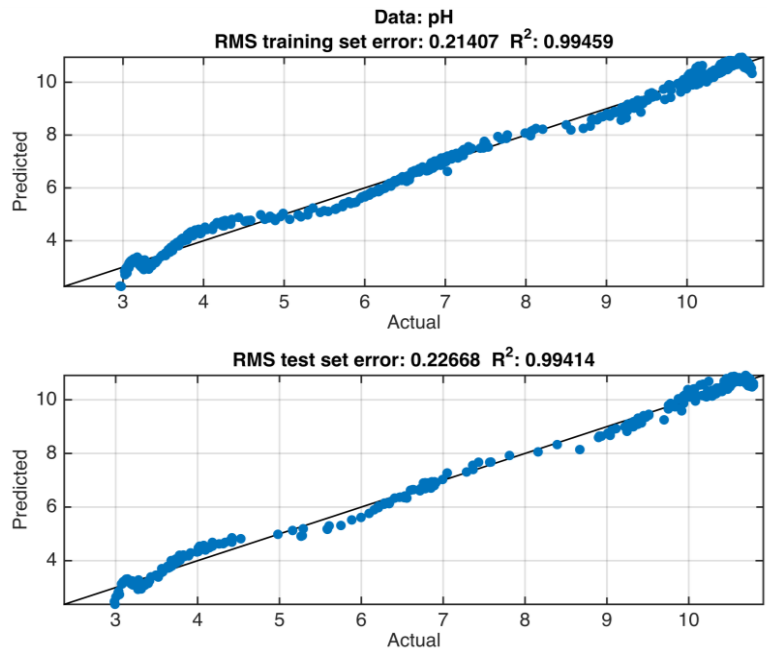

**Fig. 8.** Performance scatterplots on training and testing data sets for a selected multigene regression model. Generated by the `runtree` function.

Additionally, for any model a standalone HTML report containing detailed tabulated run configuration, performance and structural (simplified model equations and trees structures) data may be generated using the `gpmodelreport` function. These reports contain interactive scatter charts similar to that in Fig. 8. The reports are fairly lengthy – however – and so are not illustrated here.

A way of comparing the performance of a small set of models simultaneously is to generate regression error characteristic (REC; [16]) curves using the `compareModelsREC` function. REC curves are similar to receiver operating characteristic curves (ROC) used to graphically depict the performance of classifiers on a data set. An example of REC curves generated using the `compareModelsREC` function is shown below in Fig. 9. The user can specify what curves to compare in the arguments to the function, e.g.

```
compareModelsREC(gp,[2 3 9], true)
```

where the final Boolean `true` argument indicates that the best model on the training data should also be plotted in addition to models 2, 3 and 9.

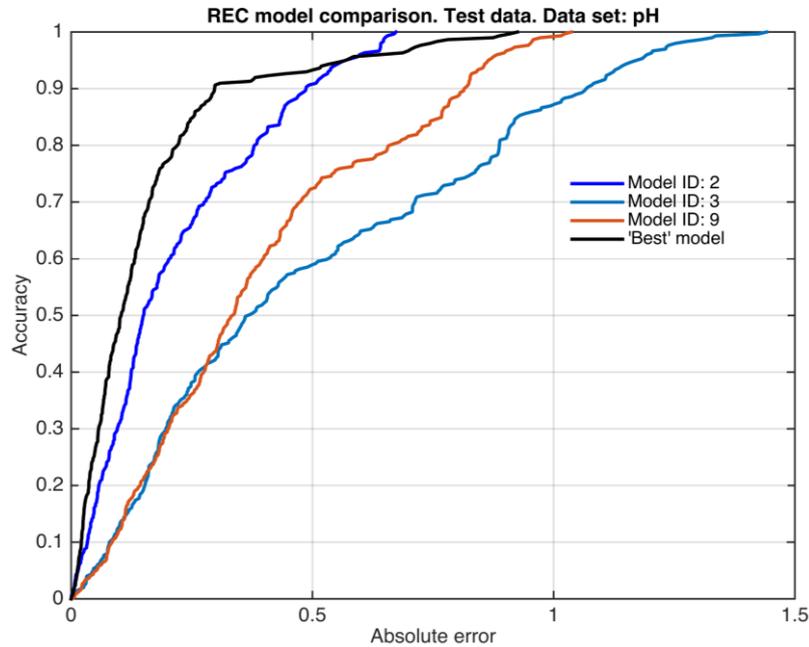

**Fig. 9.** Regression error characteristic (REC) curves. GPTIPS 2 allows the simple comparison between multigene regression models in terms of REC curves which are similar to receiver operating characteristic (ROC) curves for classifiers. The REC curves show the proportion of data points predicted ($y$ axis) with an accuracy better than the corresponding point on the $x$ axis. Hence, 'better' models lie to the upper left of the diagram. Generated using the `compareModelsREC` function.

### *4.4 Model Conversion and Export*

Finally, there is a variety of functions provided to convert and/or export models to different formats, e.g. to convert a model with numeric ID 5 to a standalone MATLAB M file called `model.m` then the `gpmodel2mfile` function may be used as follows:

```
gpmodel2mfile(gp,5,'model')
```

To convert a model to a symbolic math object, the `gpmodel2sym` function may be used in a similar way. A symbolic math object can then be converted to a string containing a snippet of C code using the `ccode` function.

# 5 Reducing Model Complexity using Gene Analysis

## 5.1 Horizontal Model Bloat

GP frequently suffers from the phenomenon of 'bloat', i.e. the tendency to evolve trees that contain terms that confer little or no performance benefit, e.g. see [19]. In terms of model development this is related to the phenomenon of overfitting. GPTIPS 2 contains a number of mechanisms intended to mitigate this. For instance: the use of fairly stringent restrictions on maximum tree depth (to ameliorate vertical bloat), the use of tree expressional complexity as a measure of model complexity (rather than a simple node count) to promote flatter trees over deeper ones during the simulated evolutionary process, the integration of the train-validate-test model development cycle, and the use of Pareto tournaments to select models that perform well (in terms of goodness of fit) and are not overly complex.

However, the use of multigene regression models in GPTIPS leads to another type of bloat that is referred to here as horizontal bloat. This is the tendency of multigene models to acquire genes that are either performance neutral (i.e. deliver no improvement in $R^2$ on the training data) or offer very small incremental performance improvements. Clearly - in the majority of practical applications - these terms are undesirable.

Horizontal bloat is the essentially the same behaviour exhibited by non-regularised MLR models, where it is well known that the addition of model terms leads to a monotonically increasing $R^2$ on training data even though the terms may not be meaningful (e.g. they are capturing noise) or allow the model to generalise well to testing or validation data sets. Multigene regression is a type of pseudo-linear MLR model and it suffers from the same problem. A typical way to combat this behaviour in MLR is to employ a method of regularisation to penalise for model complexity (e.g. ridge regression [20] and the lasso [21]). These methods can be difficult to tune in practice, however.

Ostensibly, the simplest way to way to prevent horizontal bloat in multigene regression is to limit the maximum allowed number of genes $G_{max}$ in a model. In practice, however, it is not usually easy to judge the optimal value of $G_{max}$ for any given problem. An alternative approach - and one that emphasises the human factor in instilling trust in models - is to provide a software mechanism that guides the user to take high performance models and delete selected genes to reduce the model complexity whilst maintaining a relatively high goodness of fit in terms of $R^2$. In the following section GPTIPS 2 functionality for expediting this process is described.

## *5.2 Unique Gene Analysis*

In GPTIPS 2, a new way of analysing the unique genes contained in a population of evolved models has been developed. This allows the user to visualise the genes in a population and to identify genes in an existing model that can be removed thus reducing model complexity whilst having only a relatively small impact on the model's predictive performance. The visualisation aspect (i.e. the ability to see the gene equation and the $R^2$ value if the gene were removed) is important because it allows the user to rapidly make an informed choice about which model terms to remove. Often this choice is based on problem domain knowledge of the system being modelled. For example, the user might want to delete a model term such as $\sin(1 - x^3)$ because it is inconsistent with his or her knowledge about the underlying data or system. This gene-centric visualisation allows users to tailor evolved models to suit their own preferences and knowledge of the modelled data.

An additional benefit of being able to visualise the genes in a model is that it expedites the process of human understanding of the model and intuition into which model terms account for a high degree of predictive ability and which account for lower amounts.

After a GPTIPS run has been completed, the user can extract a MATLAB data structure containing all of the unique genes in a population using the `uniquegenes` function as indicated below:

```
genes = uniquegenes(gp)
```

This function does the following:

- Extracts every genotype i.e. tree encoded gene (gene weights are ignored) from each model in the population.
- Deletes duplicate genotypes.
- Converts the unique genotypes to symbolic math objects (phenotypes) and then analytically simplifies them using MATLAB's symbolic math engine (MuPAD).
- Deletes any duplicate symbolic math objects representing genes and assigns a numeric ID to the remaining unique gene objects.

Note that it is quite frequent that two different genotypes will, after conversion to symbolic math objects and automated analytic simplification, resolve to the same phenotype.

Next - to provide an interactive visualisation of the genes in the population and a selected model - the `genebrowser` function is used. In the example below, it is used on the model that performed best (in terms of $R^2$) on the training data.

```
genebrowser(gp,genes,'best')
```

Clicking on any blue bar shows a yellow popup containing the symbolic version of the gene and the reduction in $R^2$ that would result if that gene were to be removed from the model. Conversely, clicking on any orange bar in the lower axis does the same for genes that are not in the current model and shows the increase in $R^2$ that would be attained if that gene were added to the model.

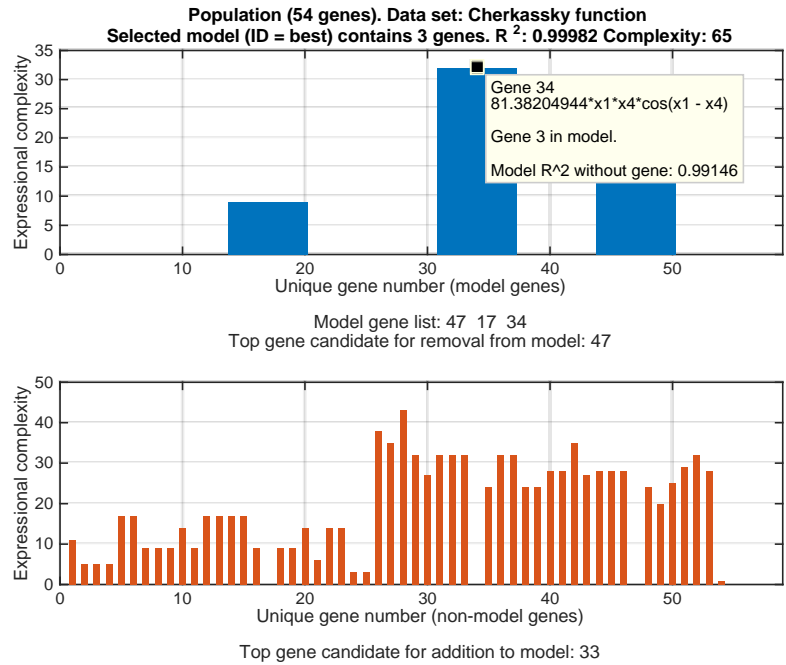

**Fig. 6.** Reducing model complexity using the `genebrowser` analysis tool. The upper bar chart shows the gene number and expressional complexity of genes comprising the selected model. The lower bar chart shows genes in the population but not in the selected model. Clicking on a blue bar representing a model gene reveals a popup containing the gene equation and the $R^2$ (on the training data) if that gene were removed from the model. Here it shows that the highlighted gene/model term $81.382\,x_1\,x_4\cos(x_1-x_4)$ is a horizontal bloat term and could be removed from the model with a very minor decrease in $R^2$.

Once the user has identified a suitable gene to be removed from the model, a new model without the gene can be generated using the `genes2gpmodel` function using the unique gene IDs as input arguments. The data structure returned from this function can be examined using the provided tools - as well as exported in various formats - in exactly the same way as any model contained within the population.

# 6 Conclusions

In this chapter GPTIPS 2, the latest version of the free open source software platform for symbolic data mining, has been described. It is emphasised that the software is aimed at non-experts in machine learning and computing science – and that the software tools provided within GPTIPS are intended to facilitate the discovery, understanding and deployment of simple, useful symbolic mathematical models automatically generated from non-linear and high dimensional data.

In addition, it has been emphasised that GPTIPS is also intended as an enabling technology platform for researchers who wish to add their own code in order to investigate symbolic data mining problems such as symbolic classification and symbolic optimisation. Whilst this article has focused largely on symbolic regression, future updates to GPTIPS 2 will include improved out-of-the-box functionality to support symbolic classification.

Finally, it is noted that GPTIPS 2 provides a novel gene-centric approach (and corresponding visual analytic tools) to identifying and removing unnecessary complexity (horizontal bloat) in multigene regression models, leading to the identification of accurate, user tailored, compact and data driven symbolic models.

# References

1. Koza J.R. (1992) Genetic programming: on the programming of computers by means of natural selection, The MIT Press, Cambridge (MA).
2. Espejo, P.G., Ventura, S., Herrera, F. (2010) A survey on the application of genetic programming to classification, IEEE Transactions on Systems, Man and Cybernetics - Part C: Applications and Reviews, 40 (2), 121--144.
3. Morrison, G., Searson, D., Willis, M. (2010) Using genetic programming to evolve a team of data classifiers. World Academy of Science, Engineering and Technology, International Science Index 48, 4(12), 210--213.
4. Pan, I., Das, S. (2014) When Darwin meets Lorenz: Evolving new chaotic attractors through genetic programming. arXiv preprint arXiv:1409.7842.
5. Gandomi, A.H., Alavi, A.H. (2011) A new multi-gene genetic programming approach to non-linear system modeling. Part II: geotechnical and earthquake engineering problems, Neural Comput & Applic, 21(1), 171--187.
6. Smits, G.F., Kotanchek, M. (2004) Pareto-front exploitation in symbolic regression, Genetic Programming Theory and Practice II, 283--299.
7. Poli, R., Langdon, W.B., McPhee, N.F., Koza, J.R. (2007). Genetic programming: An introductory tutorial and a survey of techniques and applications. University of Essex, UK, Tech. Rep. CES-475.
8. Pan, I., Pandey, D.S., Das, S. (2013) Global solar irradiation prediction using a multi-gene genetic programming approach. Journal of Renewable and Sustainable Energy, 5(6), 063129.
9. Barati, R., Neyshabouri, S.A.A.S., Ahmadi, G. (2014) Development of empirical models with high accuracy for estimation of drag coefficient of flow around a smooth sphere: An evolutionary approach. Powder Technology, 257, 11--19.
10. Floares, A.G., Luludachi, I. (2014) Inferring transcription networks from data. Springer Handbook of Bio-/Neuroinformatics, Springer Berlin Heidelberg, 311--326.

11. Gandomi, A.H., Alavi, A.H. (2012) A new multi-gene genetic programming approach to nonlinear system modeling. Part I: materials and structural engineering problems. Neural Computing and Applications, 21(1), 171--187.
12. Searson, D.P. (2002) Non-linear PLS using genetic programming, PhD thesis, Newcastle University, UK.
13. Searson D.P., Willis M.J., Montague, G.A. (2007) Co-evolution of non-linear PLS model components, Journal of Chemometrics, 21 (12), 592--603.
14. Searson, D.P., Leahy, D.E., Willis, M.J. (2010) GPTIPS: an open source genetic programming toolbox for multigene symbolic regression, Proceedings of the International MultiConference of Engineers and Computer Scientists 2010 (IMECS 2010), Hong Kong, 17-19 March.
15. Deb, K., Pratap, A., Agarwal, S., Meyarivan, T.A.M.T (2002) A fast and elitist multiobjective genetic algorithm: NSGA-II. Evolutionary Computation, IEEE Transactions on, 6(2), 182--197.
16. Bi, J., Bennett, K.P. (2003) Regression error characteristic curves, Proceedings of the Twentieth International Conference on Machine Learning (ICML-2003), Washington DC, 43--50..
17. Keijzer, M. (2004) Scaled symbolic regression, Genetic Programming and Evolvable Machines, 5, 259--269.
18. Storn, R., Price, K. (1997) Differential evolution – a simple and efficient heuristic for global optimization over continuous spaces. Journal of global optimization, 11(4), 341--359.
19. Luke, S., Panait, L. (2006) A comparison of bloat control methods for genetic programming, Evol. Comput., 14(3), 309--344.
20. Hoerl, A. E., Kennard, R.W. (1970) Ridge regression: Biased estimation for nonorthogonal problems. Technometrics, 12(1), 55--67.
21. Tibshirani, R. (1996) Regression shrinkage and selection via the lasso. Journal of the Royal Statistical Society. Series B (Methodological), 267--288.